\documentclass[twocolumn]{aa} 
\usepackage{lineno}
\usepackage{graphicx}
\usepackage{graphics}

\usepackage{longtable}
\usepackage{txfonts}

\usepackage{epsf,epsfig}
\usepackage[dvips]{color}
\usepackage{color}
\usepackage{amssymb,amsfonts}
\usepackage{latexsym}

\usepackage{tabularx}
\usepackage{multirow}

\usepackage{natbib}
\bibpunct{(}{)}{;}{a}{}{,} 

\begin{document}



\newcommand{\vla}{VLA}
\newcommand{\most}{MOST}
\newcommand{\park}{Parkes}
\newcommand{\atc}{ATCA}
\newcommand{\rosat}{{\it ROSAT}}
\newcommand{\rxte}{{\it RXTE}}
\newcommand{\xmm}{{\it XMM-Newton}}
\newcommand{\chan}{{\it Chandra}}
\newcommand{\sax}{{\it BeppoSAX}}
\newcommand{\asca}{{\it ASCA}}
\newcommand{\suz}{{\it Suzaku}}
\newcommand{\fermi}{\emph{Fermi}}
\newcommand{\hess}{H.E.S.S.}



\newcommand{\un}[1]{~\hspace{-1pt}\ensuremath{\mathrm{#1}}}
\newcommand{\am}{$^{\prime}$}
\newcommand{\as}{$^{\prime\prime}$}
\newcommand{\gammaray}{$\gamma$-ray}
\newcommand{\gammarays}{$\gamma$-rays}
\newcommand{\xray}{X-ray}
\newcommand{\xrays}{X-rays}

\newcommand{\rxj}{RX~J1713.7$-$3946}
\newcommand{\rcw}{RCW~86}
\newcommand{\msh}{MSH~14$-$6\textit{3}}


\def\neff{$\bar{n}$}
\def\Veff{$\bar{V}$}
\def\ks{km s$^{-1}$}
\def\kms{$\mathrm{km}$ $\mathrm{s}^{-1}$}
\def\d{$^\circ$}
\def\m{$^\prime$}
\def\s{$^{\prime\prime}$}
\def\hh{$^{\mathrm h}$}
\def\mm{$^{\mathrm m}$}
\def\second{$^{\mathrm s}$}
\def\cm3{cm$^{-3}$}
\def\pp{$^{\prime\prime}$}
\def\msun{M$_\odot$}

\def\eg{{\it e.g.~}}
\def\etal{et~al.~}
\def\ie{{\em i.e.~}}

\def\res#1#2#3{$#1_{-#2}^{+#3}$}


\title{Constraints on cosmic-ray efficiency in the supernova remnant \\
       \rcw~using multi-wavelength observations}

\author{M.~Lemoine-Goumard$^{1,2}$ \and M.~Renaud$^{3}$ \and J.~Vink$^{4}$ \and G.~E.~Allen$^{5}$ \and A.~Bamba$^{6}$ \and F.~Giordano$^{7,8}$ \and Y.~Uchiyama$^{9}$}

\authorrunning{Lemoine-Goumard et al.}
\offprints{lemoine@cenbg.in2p3.fr, mrenaud@lupm.univ-montp2.fr}

\institute{
\inst{1}~Universit\'e Bordeaux 1, CNRS/IN2P3, Centre d'\'Etudes Nucl\'eaires de Bordeaux Gradignan, 33175 Gradignan, France \\
\email{lemoine@cenbg.in2p3.fr} \\
\inst{2}~Funded by contract ERC-StG-259391 from the European Community \\
\inst{3}~Laboratoire Univers et Particules de Montpellier, Universit\'e Montpellier 2, CNRS/IN2P3, 34095 Montpellier Cedex 5, France \\ 
\email{mrenaud@lupm.univ-montp2.fr} \\
\inst{4}~Astronomical Institute, Utrecht University, P.O. Box 80000, 3508TA Utrecht, The Netherlands \\
\inst{5}~MIT Kavli Institute for Astrophysics and Space Research, Cambridge, MA 02139, USA \\
\inst{6}~Department of Physics and Mathematics, Aoyama Gakuin University 5-10-1 Fuchinobe, Chuo-ku, Sagamihara, Kanagawa, 252-5258, Japan \\
\inst{7}~Dipartimento di Fisica "M. Merlin" dell'Universit\'a e del Politecnico di Bari, I-70126 Bari, Italy \\
\inst{8}~Istituto Nazionale di Fisica Nucleare, Sezione di Bari, 70126 Bari, Italy \\
\inst{9}~W.~W.~Hansen Experimental Physics Laboratory, Kavli Institute for Particle Astrophysics and Cosmology, Department of Physics and
SLAC National Accelerator Laboratory, Stanford University, Stanford, CA 94305, USA
}


\date{Accepted}

\abstract
{Several young supernova remnants (SNRs) have recently been detected in the high-energy (HE; 0.1 $<$ E $<$ 100 GeV) and very-high-energy (VHE; E $>$ 100 GeV) gamma-ray domains. As exemplified by \rxj, the nature of this emission has been hotly debated, and 
  direct evidence for the efficient acceleration of cosmic-ray protons at the SNR shocks still remains elusive.}
{We study the broadband gamma-ray emission from one of these young SNRs, namely \rcw, for which several observational lines of 
  evidence indirectly point towards the presence of accelerated hadrons. We then attempt to detect any putative hadronic signal from 
  this SNR in the available gamma-ray data, in order to assess the level of acceleration efficiency.}
{We analyzed more than 40 months of data acquired by the Large Area Telescope (LAT) on-board the {\emph{Fermi Gamma-Ray Space Telescope}} in the HE domain, and gathered all
  of the relevant multi-wavelength (from radio to VHE gamma-rays) information about the broadband nonthermal emission from \rcw. For this purpose, we re-analyzed the archival \xray~data from the \asca/Gas Imaging Spectrometer (GIS), the \xmm/EPIC-MOS, and the 
  \rxte/Proportional Counter Array (PCA).}
{Beyond the expected Galactic diffuse background, no significant gamma-ray emission in the direction of \rcw~is detected in any of the 0.1--1, 1--10 and 10--100 GeV \fermi-LAT maps. The derived HE upper limits, together with the \hess~measurements in the VHE
  domain, are incompatible with a standard E$_{p}^{-2}$ hadronic emission arising from proton-proton interactions, and can only be accommodated by a spectral index $\Gamma \leq$ 1.8, i.e. a value in-between the standard (test-particle) index and the asymptotic
  limit of theoretical particle spectra in the case of strongly modified shocks. In such a hadronic scenario, the total energy in accelerated 
  particles is at the level of $\eta_{CR}$ = E$_{{\rm CR}}$/E$_{{\rm SN}}$ $\sim$ 0.07 d$^{2}_{{\rm 2.5kpc}}$/\neff$_{{\rm cm-3}}$ (with
  the distance d$_{\rm 2.5kpc}$ $\equiv$ d/2.5 kpc and the effective density \neff$_{{\rm cm-3}}$ $\equiv$ \neff/1 cm$^{-3}$), and the
  average magnetic field must be stronger than 50 $\mu$G in order to significantly suppress any leptonic contribution. On the other hand, the
  interpretation of the gamma-ray emission by inverse Compton scattering of high energy electrons reproduces the multi-wavelength data
  using a reasonable value for the average magnetic field of 15--25 $\mu$G. In this leptonic scenario, we derive a conservative upper limit 
  to $\eta_{CR}$ of 0.04 d$^{2}_{{\rm 2.5kpc}}$/\neff$_{{\rm cm-3}}$. We discuss these results in the light of existing estimates of the magnetic field strength, the effective density and the acceleration efficiency in \rcw.}
{}

\titlerunning{The SNR \rcw~using multi-wavelength observations}

\maketitle


\section{Introduction}
\label{intro}

Supernova remnants (SNRs) are thought to be the primary sources of the bulk of Galactic cosmic-ray (CR) protons observed at Earth, 
up to the knee energy at $\sim$ 3~PeV. This paradigm mainly relies on the need to have sufficiently energetic sources that could provide
the necessary power to maintain the Galactic CR energy density \citep[\eg][]{fields01}, and on the dominance of SNRs among sources of
non-thermal radio emission. Our understanding of CR acceleration in SNRs mainly relies on the so-called Diffusive Shock Acceleration theory 
\cite[in its non-linear version NLDSA, see][]{malkov01}, which is commonly invoked to explain several observational (though, indirect) lines 
of evidence for efficient particle acceleration at the SNR forward shocks up to very high energies. Among the requisites that must 
be fulfilled by this theory \cite[see][for a recent review]{blasi10}, the observed broadband nonthermal emission of individual SNRs 
is of particular interest. This emission, arising from accelerated particles that emit photons through several channels (synchrotron 
[SC], inverse-Compton [IC], nonthermal bremsstrahlung, proton-proton interactions and subsequent $\pi^0$~decay), offers new 
insights into the particle acceleration mechanisms at work in these sources, given the large number of multi-wavelength observations 
available nowadays towards many Galactic SNRs.

In particular, recent observations of young SNRs in the high-energy (HE; 0.1 $<$ E $<$ 100 GeV) and very-high-energy (VHE; E $>$ 100 
GeV) gamma-ray domains have raised several questions and triggered numerous theoretical investigations 
\citep[\eg][]{za10,berezhko10,fang11,caprioli11}. The critical issue regarding the nature of the observed gamma-ray emission 
has been intensively discussed in the literature in recent years \citep[][and references therein]{ellison10}. Nevertheless, 
these joint HE/VHE observations can help us to discriminate spectrally between the leptonic (through IC emission) and hadronic 
(through $\pi^0 \rightarrow $2$\gamma$) scenarios, as shown by the \fermi-LAT \citep{abdo11a} and \hess~\citep{hess07} observations 
towards the TeV-bright SNR \rxj, which tend to support a leptonic model for the observed gamma-ray emission \citep{ellison12,li11}. 

\rcw~\citep{rodgers60}, also known as G315.4$-$2.3 or \msh~\citep{mills61}, is a 42\m~diameter Galactic SNR in the southern sky. 
There has been much controversy about the nature of the supernova (SN) progenitor, and the SNR age and distance. This has resulted from the 
difficulties in reconciling the young age of \rcw, based on its connection with the first historical SN ever recorded in AD~185, with its 
large size, given the relatively large distance estimates, which place \rcw~at 2.3--2.8 kpc \citep{rosado96,sollerman03} near an OB 
stellar association. \citet{williams11} recently reviewed all the arguments about the nature of the SN progenitor and critically examined the available observations of \rcw. They suggest that the SNR is likely the remnant of a type Ia SN, and, from hydrodynamic 
simulations, that the off-center explosion occurred in a low-density cavity carved by the progenitor system \citep[see also][]{vink97}. 
This scenario allows one to explain at once the young age and the large distance of \rcw, which we scale in terms of d$_{2.5}$ $\equiv$ 
d/(2.5 kpc) throughout the paper.

The general outline of its nearly circular shell is similar in the radio \citep{whiteoak96,dickel01}, infrared \citep[IR,][]{williams11}, 
optical \citep[\eg][]{smith97}, and \xray~\citep{vink97,bamba00,bocchino00,borkowski01b,rho02,vink06} domains, although significant 
fine-scale differences have been reported \cite[\eg][ and Fig. \ref{fig2}]{rho02}. \xray~observations towards \rcw~have revealed the 
presence of both thermal and nonthermal emission, with very distinct morphologies: while soft \xrays~correlate with optical emission from 
non-radiative shocks and IR emission from collisionally heated dust, the continuum hard \xray~emission is seen at the edges of radio 
emission. The high-temperature plasma, which mostly contains the strong Fe K$_{\alpha}$ line emission, also shows a particular morphology 
extending towards the SNR interior, as revealed with \suz, and is thought to originate from Fe-rich ejecta heated by the reverse shock 
\citep{ueno07,yamaguchi08}. The global distribution of the Fe-rich plasma in the entire SNR has recently been mapped with 
\suz~\citep{yamaguchi11}, and the total Fe mass of $\sim$ 1 \msun~deduced from these observations, together with a relatively low ejecta
mass of 1--2 \msun, strengthens the scenario of a type Ia SN at the origin of \rcw.

Physical conditions (shock speed, ambient density, and magnetic field) vary greatly along the shell-like structure. In particular, slow shocks 
\citep[$\sim$ 600--800 km s$^{-1}$, see][]{long90,ghavamian01} and relatively high post-shock densities 
\citep[$\sim$ 2 cm$^{-3}$, see][]{williams11} have been measured in the southwest (SW) and northwest (NW) regions, while the northeast (NE) region exhibits much faster shocks \citep[$\gtrsim$ 2700 km s$^{-1}$ and 6000 $\pm$ 2800 km s$^{-1}$, see][]{vink06,helder09} and lower densities \citep[$\sim$ 0.1--0.3 cm$^{-3}$, \eg][]{yamaguchi08}. In this region, \citet{helder09} argued that at least 50\% 
of the total pressure is induced by CRs, based on the discrepancy between the measured shock velocity and the spectroscopically
determined post-shock proton temperature \citep[see also][]{vink10}. Moreover, \citet{vink06} found evidence for a concave electron 
spectrum, as predicted by the NLDSA theory, in order to explain the radio and \xray~SC emission observed from the same region. These two 
measurements, together with the recent detection of \rcw~with the \hess~experiment in the VHE domain \citep{hess09}, seem to point towards 
an efficient CR source. However, complementary HE observations are needed to probe the nature of the gamma-ray emission, as 
discussed above. 

We here report on \fermi-LAT observations and data analysis towards \rcw~(section \ref{lat}). In an attempt to constrain the broadband 
nonthermal emission from the SNR, we present a re-analysis of the archival \asca/GIS, \xmm/EPIC-MOS, and \rxte/PCA \xray~data (section
\ref{xray}), and collect the available information in the radio and VHE gamma-ray domains. We then discuss the constraints derived on the
parent particle spectrum and on the subsequent acceleration efficiency, in the light of existing estimates (section \ref{discu}).


\section{\fermi-LAT observations and data analysis}
\label{lat}

\subsection{Observations}
\label{obs}

The LAT is a gamma-ray telescope that detects photons by conversion into electron-positron pairs in the energy range from 
20~MeV to more than 300~GeV, as described by~\citet{atwood09}. It is made of a high-resolution converter/tracker (for direction measurement 
of the incident gamma-rays), a CsI(Tl) crystal calorimeter (for energy measurement) and an anti-coincidence detector to reject the 
background of charged particles. The LAT has a large effective area ($\sim$ 8000 cm$^{2}$ on-axis above 1~GeV), 
a wide field of view ($\sim$ 2.4 sr) and good angular resolution ($\sim$0.6$^{\circ}$ radius for 68$\%$ containment at 1 GeV for events converting in the front section of the tracker). The on-orbit calibration is described in \citet{abdo09a}.

The analysis used more than 40 months of data collected starting on August 4, 2008, and extending until February 22, 2012. We selected events 
with energies greater than 0.1~GeV, and excluded those with zenith angles larger than 100\d, to minimize contamination from 
secondary gamma rays from the Earth's atmosphere~\citep{abdo09b}. In addition, we excluded the time intervals when the rocking angle was more than 52\d~and when the \fermi~satellite was within the South Atlantic Anomaly. We used the P7SOURCE\_V6 Instrument Response Functions (IRFs), and selected the `Source' events, which correspond to the best compromise between the number of selected photons and the charged particle residual 
background for the study of point-like or slightly extended sources. 

\subsection{Analysis and results}
\label{ana}

For the spectral analysis, a spectral-spatial model containing diffuse and point-like sources was created, and the parameters were 
obtained from a maximum likelihood fit to the data using the \emph{gtlike} code implemented in the \fermi~Science 
Tools\footnote{\label{fssc}\texttt{\tiny http://fermi.gsfc.nasa.gov/ssc/data/}}. The likelihood tool \emph{gtlike} fits a source model to the data along with models for the residual charged particles and diffuse gamma-ray emission. For the Galactic diffuse emission, we used the ring-hybrid model {\it gal\_2yearp7v6\_v0.fits}. The 
instrumental background and the extragalactic radiation are described by a single isotropic component with the spectral shape in the 
tabulated model {\it iso\_p7v6source.txt}. These models are available from the \fermi~Science Support Center\footnote{\texttt{\tiny http://fermi.gsfc.nasa.gov/ssc/data/access/lat/BackgroundModels.html}}. All sources within 15\d~of \rcw~were extracted from the \fermi-LAT 2FGL catalog \citep{nolan12} and added to our spectral-spatial model of the 
region. The spectral parameters of sources closer than 3$^{\circ}$ to RCW~86 were allowed to vary in the likelihood fit, while the parameters of 
all other sources were fixed at the 2FGL values. 

\begin{figure}[!htb]
\begin{center}
\includegraphics[width=0.99\linewidth]{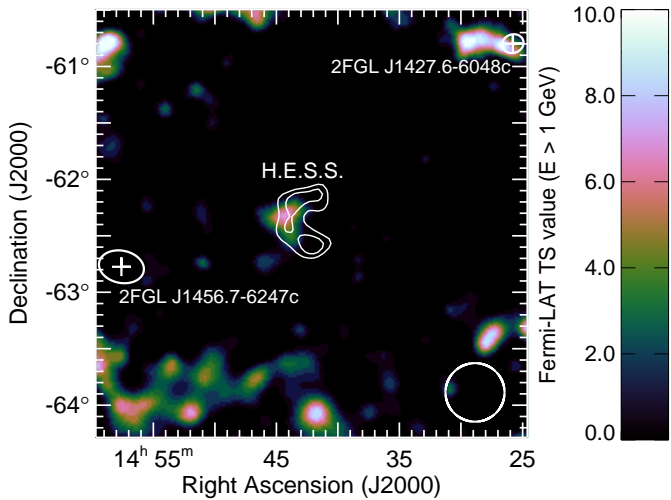}

\caption{\fermi-LAT test statistic (TS) map towards \rcw, above 1~GeV. The TS was evaluated by placing a point-source at the center of each pixel, Galactic diffuse emission and nearby 2FGL sources being included in the background model. 
         \hess~contours \citep{hess09} are overlaid 
         as white solid lines, and the white circle indicates the size of the 68\% containment region of the LAT PSF at 10 GeV.
         Crosses and ellipses mark the best-fit positions and associated errors of the two sources from the 2FGL
         \fermi-LAT catalog \citep{nolan12} within 4\d~from \rcw. \label{fig1}}

\end{center}
\end{figure}

Using this spectral-spatial model, we computed the LAT test statistic (TS) map of the region containing \rcw, as shown in Figure~\ref{fig1}. The TS is defined as twice the difference between the log-likelihood $L_1$ obtained by fitting 
a source model plus the background model to the data, and the log-likelihood $L_0$ obtained by fitting the background model only, i.e. TS = 2($L_1$ - $L_0$). A TS = 25 with two degrees of freedom (spectral index and normalization) roughly corresponds to a 4.6$\sigma$ detection significance. Figure~\ref{fig1} 
contains the TS value for a point source with fixed photon index $\Gamma = 2$ at each map location, thus giving a measure of the 
statistical significance for the detection of a gamma-ray source in excess of the background. No significant excess suggesting gamma-ray 
emission is evident in the skymap, apart from faint excesses, which are unassociated with any 2FGL catalogued source, at the northern and southern 
borders of the map, and a faint hotspot located near the NE shell of \rcw. Under the assumption of a point-like source, this
excess is found at R.A.(J2000) $\approx$ 221.0\d~and Dec(J2000) $\approx$ -62.4\d~with a TS value of 12 (\ie~$\sim$ 2.4~$\sigma$ for four dof, where the spectral index is allowed to vary). Therefore, we do not consider this hotspot in the following, bearing in mind that
further \fermi-LAT observations in the coming years might provide confirmation of this indication of a signal towards \rcw.

Since the \fermi-LAT 2FGL catalog does not contain any source in a region of 1\d~around the SNR \rcw, to determine an upper limit to the gamma-ray emission from \rcw, we added an extended source consistent with the \hess~observations, \ie~a uniform disk centered 
at (R.A.(J2000) = 220.68\d, Dec.(J2000) = -62.44\d) with a radius of 0.40\d. We modeled the emission from \rcw~using a power-law spectral distribution 
with two free parameters: the flux and the spectral index. The test statistic was employed to evaluate the significance of the 
gamma-ray fluxes coming from \rcw~in three different energy bands: 0.1 GeV -- 1 GeV, 1 GeV -- 10 GeV, and 10 GeV -- 100 GeV. In all tested 
energy bands, the likelihood ratio test indicate values of TS less than 25. This agrees with the non-detection in the 2FGL catalog as well as the LAT TS map presented in Figure~\ref{fig1}. Assuming a power-law shape for the source with a fixed spectral index 
of $2$, we then derived 99.9\% confidence level (CL) upper limits to the flux in these three energy intervals using the Bayesian method proposed by~\cite{helene}. These upper limits, shown
in Figure~\ref{fig4}, amount to 1.3 $\times$ 10$^{-8}$, 1.1 $\times$ 10$^{-9}$, and 3.4 $\times$ 10$^{-10}$ ph cm$^{-2}$ s$^{-1}$, 
respectively. Varying the spectral index between 1.5 and 3 did not change significantly these estimates.

\section{X-ray observations}
\label{xray}

We seek to estimate the total nonthermal \xray~emission from \rcw~in order to constrain the underlying acceleration mechanisms, 
together with the gamma-ray observations presented above. There is a large amount of \xray~data towards the SNR, since \asca, \xmm, and
more recently \suz~have performed a nearly complete coverage \citep{bamba00,vink06,yamaguchi11}. In hard \xrays, the non-imaging 
instrument PCA onboard \rxte~detected \rcw~above 10 keV, and the original results were presented in \citet{allen99}. From our point
of view, it is worthwhile revisiting these archival \xray~data for several reasons. The \rxte/PCA data analysis presented below was carried out with the latest software release, and we investigate (and hence correct the extracted spectrum for) the contamination 
from the Galactic Ridge and the effect of the PCA off-axis response for such an extended source. For both \asca/GIS and \rxte/PCA datasets, 
we also want to constrain the SC emission with the most advanced, physically-motivated, modeling dedicated to \xray~observations
\citep{za10}, rather than either the usual power-law approximation, which is obviously inadequate in the cutoff regime, or the often-used 
\texttt{srcut} model \citep{reynolds99}, which relies on the SNR properties in the radio domain and hence makes the implicit assumption 
that radio and nonthermal \xrays~arise from the same emission region, at odds with what is observed in 
\rcw~\citep[][and Fig. \ref{fig2}]{rho02}.

\subsection{ASCA/GIS}
\label{asca}

\begin{figure}[!htb]
\begin{center}
\includegraphics[width=1.0\linewidth]{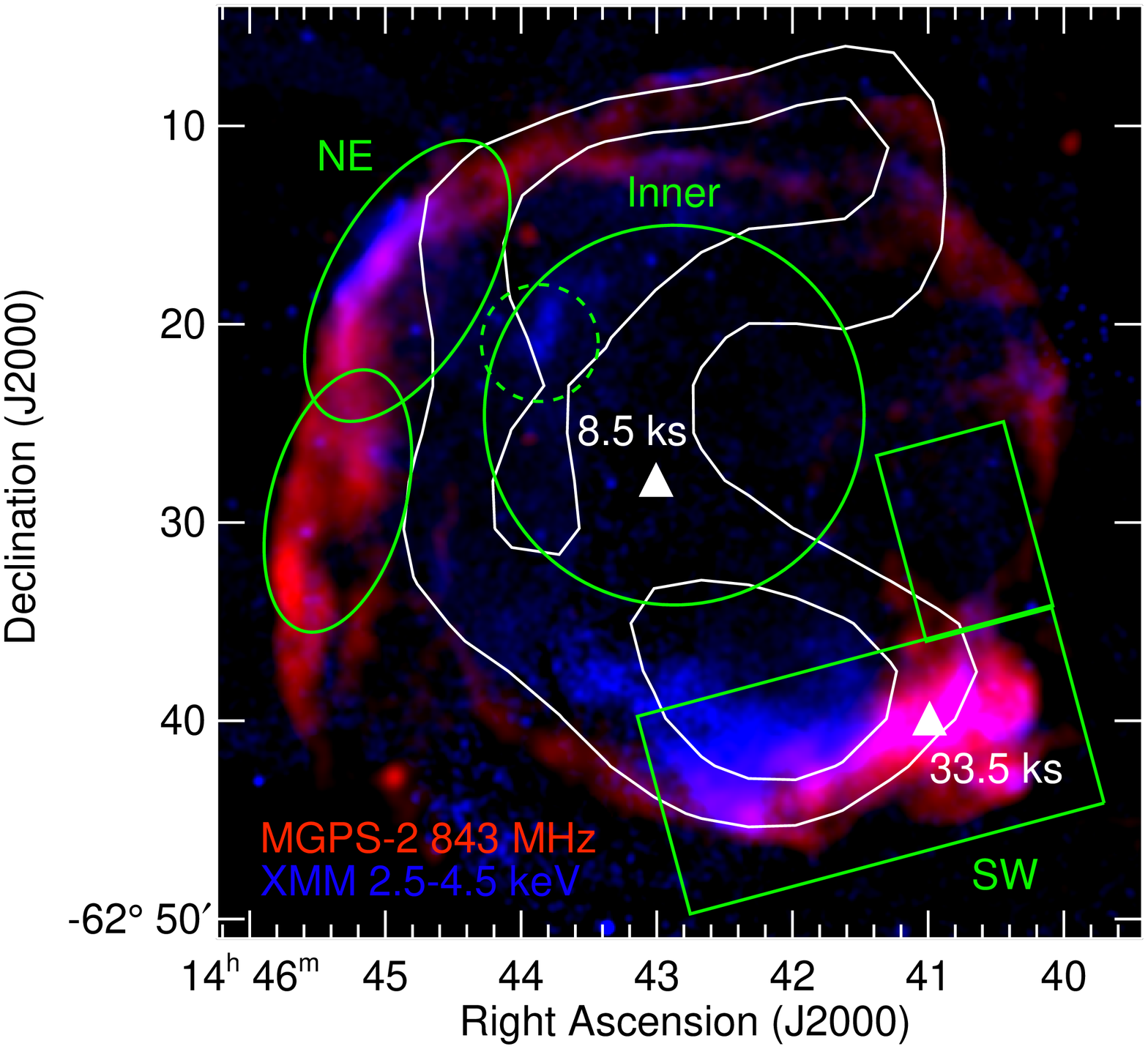}

\caption{Composite radio \citep[MGPS-2 at 843 MHz,][in red]{murphy07} and \xray~(\xmm/MOS~in the 2.5--4.5 keV band, in blue)
         image of \rcw. The \xmm/MOS image has been smoothed with a Gaussian of width 10\s. \hess~contours \citep{hess09} are 
         overlaid in white, as in Figure \ref{fig1}. Triangles mark the \rxte~pointing positions with the associated exposure
         times. The different \asca/GIS regions for spectral extraction are delineated in green and labeled NE, SW and Inner
         (counts within the dashed circle were excluded), as in \citet{bamba00}. \label{fig2}}

\end{center}
\end{figure}

\rcw~was observed with \asca~\citep{tanaka94} on 1993 August 17--18 with three pointings towards the NE, western (W), and SW regions. \asca~has two kinds of instruments, GIS \citep[Gas Imaging Spectrometers;][]{ohashi96} and SIS
\citep[Solid-state Imaging Spectrometers;][]{burke91}, each at the focus of an X-ray Telescope \citep[][]{serlemitsos95}. We focus on
the GIS data set since it has a larger field of view and effective area. Data selection and analysis were performed following the
procedure described in \citet{bamba00}. The resulting exposure times are $\sim$20~ks for NE, $\sim$9~ks for W, and $\sim$8~ks for SW. We produced \asca/GIS spectra from three distinct regions of \rcw, which we labeled SW, NE and inner, as shown in Figure \ref{fig2}.


Since \rcw~exhibits both thermal and nonthermal \xray~emission at different relative levels along the shell, we fit these spectra 
individually with an absorbed three-component model consisting of (1) a plane-parallel shock model with variable metal abundances 
\citep[with their relative abundances fixed to solar ratios, \texttt{pshock},][]{borkowski01a}, (2) an unresolved Gaussian line at fixed 
energy of 6.4~keV, and (3) the analytical approximation of the SC emission \citep[][ZA10 hereafter]{za10} from the shock-accelerated 
electron spectra derived by \citet{za07} in the SC-loss-limited regime. The first two components account for the low-temperature thermal 
emission in non-equilibrium ionization and the Fe-K line detected in several regions along the SNR shell \citep{borkowski01b,ueno07}, 
respectively. We note that this iron line is most likely produced by a high-temperature ($\gtrsim$ 5 keV) plasma 
\citep{bamba00,bocchino00,borkowski01b,rho02}, which we do not model here. However, we checked that adding such a component does not change 
significantly the best-fit results. The third component was used to describe the \xray~SC emission measured in 
\rcw~\citep[\eg][]{borkowski01b,vink06}. We followed the ZA10 prescription, according to which the SC spectrum from shock-accelerated 
electrons in the downstream region \citep[see][]{za07}, in the case of Bohm diffusion and for a distributed (turbulent) magnetic field,
is found to have the following behavior:

\begin{eqnarray}
\frac{dN(E)}{dE} & \propto & E^{-2} \left[1 + 0.185 \left(\frac{E}{E_1}\right)^{0.4}\right]^{25/4.8} \times{\mbox{exp}\left[-\left(\frac{E}{E_1}\right)^{1/3}\right]} \\
E_1 & = & \frac{\mbox{0.056 keV}}{\eta_B} \left(\frac{V_f}{3 \times{10^3 \mbox{ km s}^{-1}}}\right)^2
\label{eq1}
\end{eqnarray}

\noindent for which the upstream to downstream magnetic field ratio $\kappa$ is assumed to be 1/$\sqrt{11}$, $V_f$ is the forward shock
velocity, and $\eta_B \equiv D/D_{\mathrm{Bohm}} \geq 1$ is the factor allowing for deviation of the diffusion coefficient from its nominal 
value. The above expression is valid when the electron spectrum in the cutoff region is determined by the energy losses, \ie when the SNR 
age is older than the SC cooling time of electrons $\tau \sim$ 1300 $\eta_B^{1/2}$ ($V_f$/10$^3$ km s$^{-1}$)$^{-1}$ 
(B/25 $\mu$G)$^{-3/2}$ yr. Provided that the diffusion proceeds close to the nominal Bohm regime, this condition is met for \rcw, which 
exhibits shock speeds in excess of 700 km s$^{-1}$ and B-field estimates of at least 25 $\mu$G \citep{volk05,vink06}. 

\begin{figure}[!htb]
\begin{center}
\includegraphics[width=0.9\linewidth]{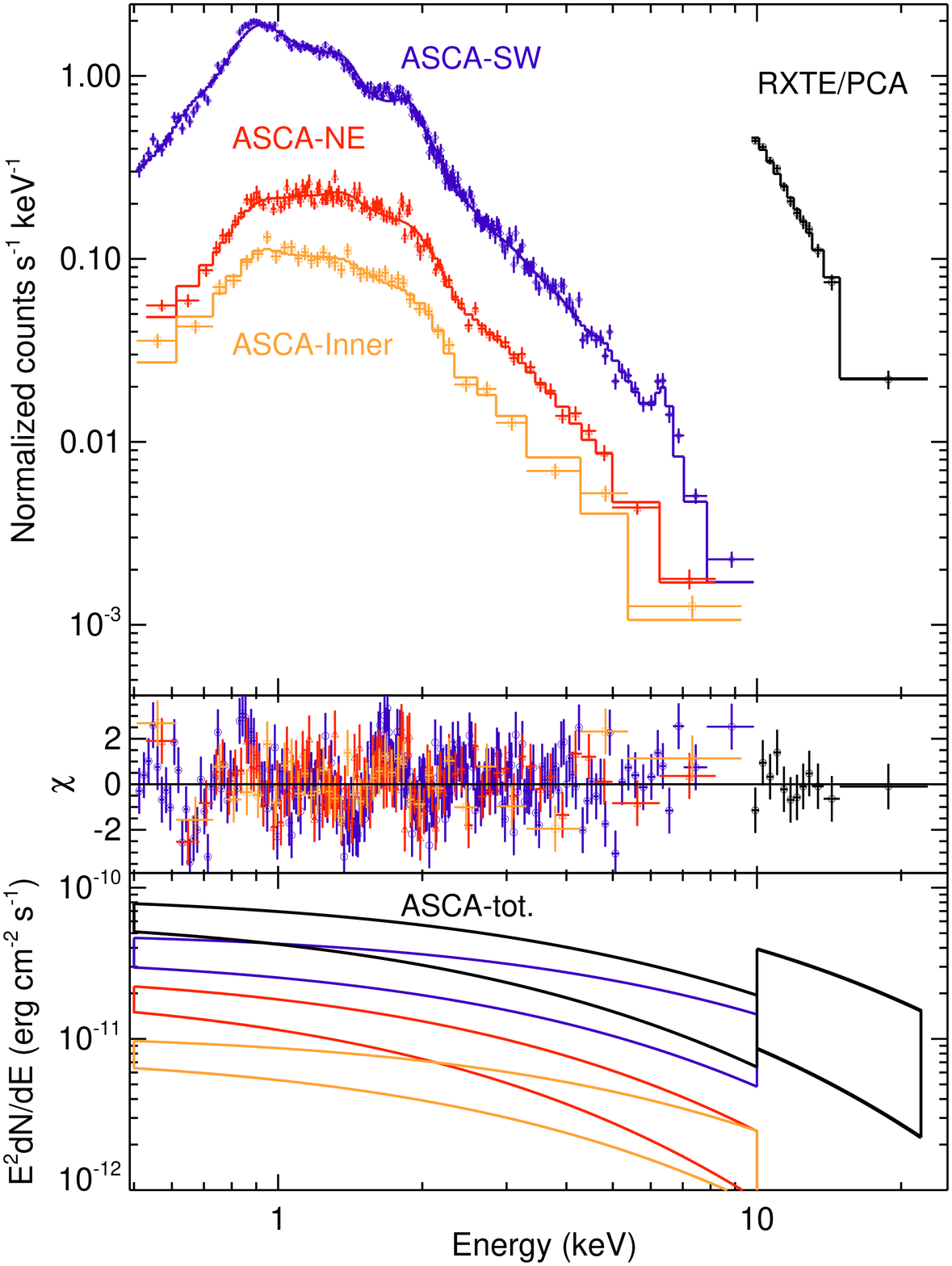}

\caption{0.5--10 keV \asca/GIS spectra from the three regions considered in this work, together with the 
         \rxte/PCA spectrum in the 10--25 keV energy range. All spectra have been rebinned to 10 $\sigma$ 
         per bin for sake of clarity. Solid lines represent the best-fit models (see text and Table \ref{tab1}). 
         Residuals and unfolded {\it nonthermal} spectra are shown in the middle and lower panels, respectively. \label{fig3}}

\end{center}
\end{figure}

The \asca/GIS~spectra and their respective best-fit models are shown in Figure \ref{fig3}, while the spectral parameters are provided in
Table \ref{tab1}. Results are in good agreement with those reported by \citet{bamba00} and \citet{borkowski01b}, who both used \asca~data 
and different spectral models. Regarding the nonthermal emission, the best-fit values of E$_1$ $\sim$ 0.01--0.02 keV translate into shock 
speeds on the order of 1300--1800 $\sqrt{\eta_B}$ km s$^{-1}$, which are roughly compatible with previous estimates for the different parts 
of the \rcw~shell. 
As shown in Table \ref{tab1}, the ionization ages $\tau$ from the \texttt{pshock} models are poorly constrained, and we fixed them to their 
nominal values while extracting the errors in the other free parameters. Finally, the three defined regions do actually not encompass the
whole SNR. To estimate the nonthermal flux outside the \asca~regions, we used the 2.5--4.5 keV \xmm/MOS image shown in Figure
\ref{fig2}, assuming that the flux in this band is entirely nonthermal. We found that 80\% of the 2.5--4.5 keV flux is contained within the 
\asca~regions, and we consider in the following the best-fit unfolded spectra shown in Figure \ref{fig3} (bottom) corrected by this
factor.

\subsection{XMM-Newton/MOS}
\label{xmm}

\xmm~observed \rcw~during several cycles, as part of the guaranteed time (GT, PI: Bleeker) program in 2000 and 2001 and as part of the
guest observer program (PI: Vink) in 2004 and 2007. The advantage of \xmm~observations over the \asca~observations is that almost the
entire SNR was covered, and that \xmm~has a larger effective area. This allowed us to estimate the total nonthermal \xray~flux, including
that from regions of faint nonthermal emission outside the \asca~regions. Unfortunately, the very elliptical and high orbit of \xmm~results 
in a higher background rate than \asca, which, moreover, is very irregular. To obtain a nonthermal flux estimate as accurately as
possible we limited ourselves to the observation IDs the least plagued by a high background signal. The data reduction was carried out 
with the standard \xmm~software SAS version 20110223. To limit the effects of chip-gaps and variable background, we limited the
analysis to the EPIC-MOS instruments. Since the \xmm~observations were spread out over a large time interval, with many different, overlapping pointings, we based our nonthermal flux estimate on imaging rather than on a full spectral analysis, since for the latter one
would need to take into account the different backgrounds per spectrum and different exposure weights for each single extraction region.
The image for the EPIC-MOS 1 and 2 instruments, shown in Figure \ref{fig2}, was made in the 2.5--4.5 keV energy band. This band was
used to estimate the total nonthermal flux, which starts dominating above $\sim 2$~keV although some S XV line emission can still be
present at 2.45 keV. 
The cut-off at 4.5 keV was used for practical reasons, as the relative background signal increases at higher energies. The effective area
between 2.5 and 4.5 keV is relatively flat and we adopt here the mean value of 345~cm$^2$. The background estimates were based on the
regions of the image outside \rcw. This is not ideal, as the background rate varies from observation to observation, with an estimated
background level of (0.75--1.5) $\times$ 10$^{-4}$~events\ s$^{-1}$ arcsec$^{-2}$. The total count rate in the 2.5--4.5 keV band for the
region containing \rcw~is 3.4 counts\ s$^{-1}$ (for a total number of events of 3.7 $\times$ 10$^5$), which reduces to 1.2--2.3 counts
s$^{-1}$ after background subtraction. To convert this count rate to nonthermal flux, we used two approaches. One was based on the analysis
of the nonthermal emission from the NE region \citep{vink06}. For the MOS spectra of this region, we have both a flux estimate
and a total 2.5--4.5 keV count rate. Using the best-fit spectral index of $\Gamma$ = 2.8, we found that the conversion factor from count
rate to nonthermal flux in the 2.5--4.5 keV band is $\sim$ 1.7 $\times$ 10$^{-11}$~erg cm$^{-2}$ s$^{-1}$/(cnt/s). Hence, the 2.5--4.5 keV
nonthermal flux amounts to (2.1--3.9) $\times$ 10$^{-11}$ erg cm$^{-2}$ s$^{-1}$, corresponding to a normalization at 1~keV of (5.7--11)
$\times$ 10$^{-2}$ cm$^{-2}$ s$^{-1}$ keV$^{-1}$. One can also directly convert the net count rate to a flux rate using the effective area
given above, which gives a normalization at 1 keV of (5.1--9.8) $\times$ 10$^{-2}$ cm$^{-2}$ s$^{-1}$ keV$^{-1}$. These values are
consistent with the measurements by \asca~(and \rxte, see below and Table \ref{tab1}).

\subsection{RXTE/PCA}
\label{rxte}

The SNR was observed with \rxte~for a total of 42~ks between 1997 February 26 and 1997 March 17. The PCA \citep{jahoda06} is comprised of 
an array of five coaligned proportional counter units (PCUs) that are mechanically collimated to have a field of view of 1\d~(FWHM). For 
\rcw, the on-axis effective area has a peak of about 4600~cm$^{2}$ at 7.3~keV and is greater than or equal to 10\% of this value from about 
2.4 to 27~keV. Between these energies, the fractional energy resolution ${\rm \Delta}E / E$ ranges from about 0.25 to 0.12 (FWHM), 
respectively. The data for observation ID 20260 were downloaded from the HEASARC archive and reprocessed using version 6.11 of
FTOOLS\footnote{\texttt{\tiny http://heasarc.nasa.gov/ftools/ftools\_menu.html}}. The tools {\tt xtefilt} and {\tt maketime} were used to
produce a new filter file for each one of the five pointings in the observation (whose locations are shown in Figure \ref{fig2}) and a set
of good-time intervals (GTIs) for standard event selections, respectively.
The GTIs, which include 38~ks of data, and the tool {\tt saextrct} were used to obtain a single, combined spectrum for the first layer of
anodes in all five PCUs.
An estimate of the combined spectrum of the instrumental and cosmic \xray~backgrounds was produced using the tool {\tt pcabackest} with the 
latest combined faint-source model and SAA history file. As a test of the accuracy of this model, the total number of background counts in
the last 50 of 129 pulse-height channels (\ie~at energies above 43~keV) is 698,135. For comparison, the total number of counts in the data
in the same range is 697,541. The two values agree to better than 0.1\% (\ie~to better than 1~$\sigma$). A response file was produced using 
the tool {\tt pcarsp}. 

\begin{table*}[!htb]
  \begin{center}
    \begin{tabular}{|l|c|c|c|c|}
      \hline\hline
      & ASCA-SW & ASCA-NE & ASCA-Inner & RXTE/PCA \\
      Model parameters & & & & \\
      \hline
      N$_{H}$ ($\times$ 10$^{21}$ cm$^{-2}$)  & \res{1.78}{0.44}{0.44} & \res{3.44}{0.70}{0.98} & \res{0.10}{0.10}{0.07} & -- \\ 
      \hline
      \texttt{pshock} & & & & \\ 
      {\it k}T (keV)                                 & \res{0.74}{0.07}{0.07} & \res{0.42}{0.06}{0.10} & \res{1.70}{1.05}{2.03} & -- \\
      $\tau$ ($\times$ 10$^{11}$ cm$^{-3}$ s, fixed) &            0.7         &             1.5        &           2.0          & -- \\
      Abundance                                      & \res{0.40}{0.12}{0.16} & \res{0.36}{0.21}{0.81} &        1 (fixed)       & -- \\ 
      Norm ($\times$ 10$^{-16}$ $\int$ n$_e$\,n$_H$\,dV\,/4$\pi$d$^{2}$ cm$^{-5}$) & \res{9.3}{3.0}{2.7} & \res{3.1}{3.0}{9.9} & \res{0.04}{0.01}{0.32} & -- \\ 
      \hline
      \texttt{gaussian} & & & & \\ 
      Line flux ($\times$ 10$^{-5}$ cm$^{-2}$ s$^{-1}$) & \res{9.9}{2.2}{2.2} & \res{2.7}{1.2}{1.2} & \res{3.3}{1.9}{1.9} & -- \\ 
      \hline
      \texttt{ZA10 synchrotron} & & & & \\ 
      Norm at 1 keV ($\times$ 10$^{-2}$ cm$^{-2}$ s$^{-1}$ keV$^{-1}$) & \res{3.15}{0.69}{0.40} & \res{1.76}{0.30}{0.18} & \res{0.65}{0.10}{0.11} & \res{6.18}{1.70}{3.60} \\ 
      E$_1$ ($\times$ 10$^{-2}$ keV)                                   & \res{2.20}{0.56}{0.94} & \res{1.01}{0.21}{0.20} & \res{1.82}{0.48}{0.67} & \res{2.29}{0.68}{0.79} \\ 
      \hline
      $\chi^2$ / $\nu$ & 575.8/416 & 193.4/149 & 176.8/168 & 14.4/29 \\ 
      \hline

    \end{tabular}
    \caption{Spectral models for \rcw. Uncertainties quoted here are provided 
             at the 90\% confidence level. \label{tab1}}
  \end{center}
\end{table*}

The spectral analysis was performed using the energy range from 10 to 25~keV.
Some caution should be used when interpreting the results of the PCA spectral analysis because: (1) There are spatial 
variations in the cosmic \xray~background.  In the 10--20 keV energy band, these variations are 0.2 counts s$^{-1}$ (1~$\sigma$) for the 
top layer of anodes in all five PCUs\footnote{\tiny http://universe.gsfc.nasa.gov/xrays/programs/rxte/pca/doc/bkg/bkg-2002}. For 
comparison, the background-subtracted emission from \rcw~is $1.1$ counts s$^{-1}$ in this energy band.  If an additional power-law 
component with a photon index of 1.3 \citep{ajello08} is added to the fit to represent excess cosmic \xray~background emission, then the 
best fit normalization of this component is essentially zero and the 90\%~upper limit to the normalization corresponds to 0.3 counts 
s$^{-1}$. (2) The background model does not include emission from the Galactic Ridge. Unfortunately, the analysis of the Galactic Ridge 
emission, as measured using the PCA \citep{valinia98}, does not extend to the two pointing locations for \rcw~(R.A.(J2000) = 220.75\d, Dec.(J2000) = -62.46\d~for the center and R.A.(J2000) = 220.25\d, Dec.(J2000) = -62.67\d~for the SW, see Figure \ref{fig2}). However, there is a PCA observation 
(ID 30267) of a nearby, more or less source-free, region of the sky (R.A.(J2000) = 214.55\d, Dec.(J2000) = -63.93\d). An analysis of the data for this 
observation reveals little evidence of Galactic Ridge emission in the 10--25~keV energy band, and we assume that we can ignore emission
from the Galactic Ridge in our analysis.
(3) While \rcw~is entirely within the field of view of the telescope for each pointing, the SNR is not a point source and the relative PCA
detection efficiency declines more or less linearly from one for an on-axis source to zero for a source that is 1\d~off axis
\citep{jahoda06}. Therefore, the amount of emission detected is less than the amount of emission that was produced by the SNR. On the basis of the 
\xmm~2.5--4.5 keV image and the \rxte~pointing directions shown in Figure \ref{fig2}, and the PCA off-axis response in the form
f = 1 - r/r$_0$ with r$_0$ = 0.965\d~\citep{jahoda06}, we estimate this reduction factor to be $\sim$ 0.8. The \rxte/PCA spectrum
is shown in Figure \ref{fig3}, and the best-fit parameters from the ZA10 prescription are provided in Table \ref{tab1}. Fitting the
spectrum with a power-law gives acceptable and consistent results with those originally reported by \citet{allen99}: $\Gamma$ = 3.27 $\pm$
0.14 and a normalization at 1 keV of \res{0.21}{0.06}{0.08} cm$^{-2}$ s$^{-1}$ keV$^{-1}$ ($\chi^2$/dof = 15/31). As for the total
nonthermal \asca~spectrum, we consider in the following the best-fit unfolded PCA spectrum shown in Figure \ref{fig3} (bottom) corrected
from this reduction factor.


\section{Discussion}
\label{discu}

To constrain the nature of the gamma-ray emission from \rcw, we have collected all the available information in the radio and VHE 
gamma-ray domains, in addition to the \fermi-LAT and \xray~data analyses presented in the previous sections. Observations towards \rcw~have 
been carried out with several radiotelescopes. The flux densities measured with MOST at 408 MHz (86 Jy) and Parkes at 5 GHz (18.2 Jy) are 
extracted from \citet{caswell75}, with respective errors of 10\% (D.~A. Green 2011, private communication). The lower limits of 22 and 28 
Jy were derived at 843 MHz \citep{whiteoak96} and 1.34 GHz \citep{dickel01}, respectively. Spectral data points in the VHE domain were obtained with the \hess~experiment \citep{hess09}. The spectrum is well described with a power-law between 0.6 and 60 TeV, with a 
photon index of 2.54 $\pm$ 0.12 and a normalization at 1 TeV of (3.72 $\pm$ 0.50) $\times$ 10$^{-12}$ cm$^{-2}$ s$^{-1}$ TeV$^{-1}$ 
(statistical errors only).

\subsection{Modeling the broadband emission of RCW 86}
\label{modeling}

\begin{figure*}
\begin{center}
\includegraphics[width=0.95\linewidth]{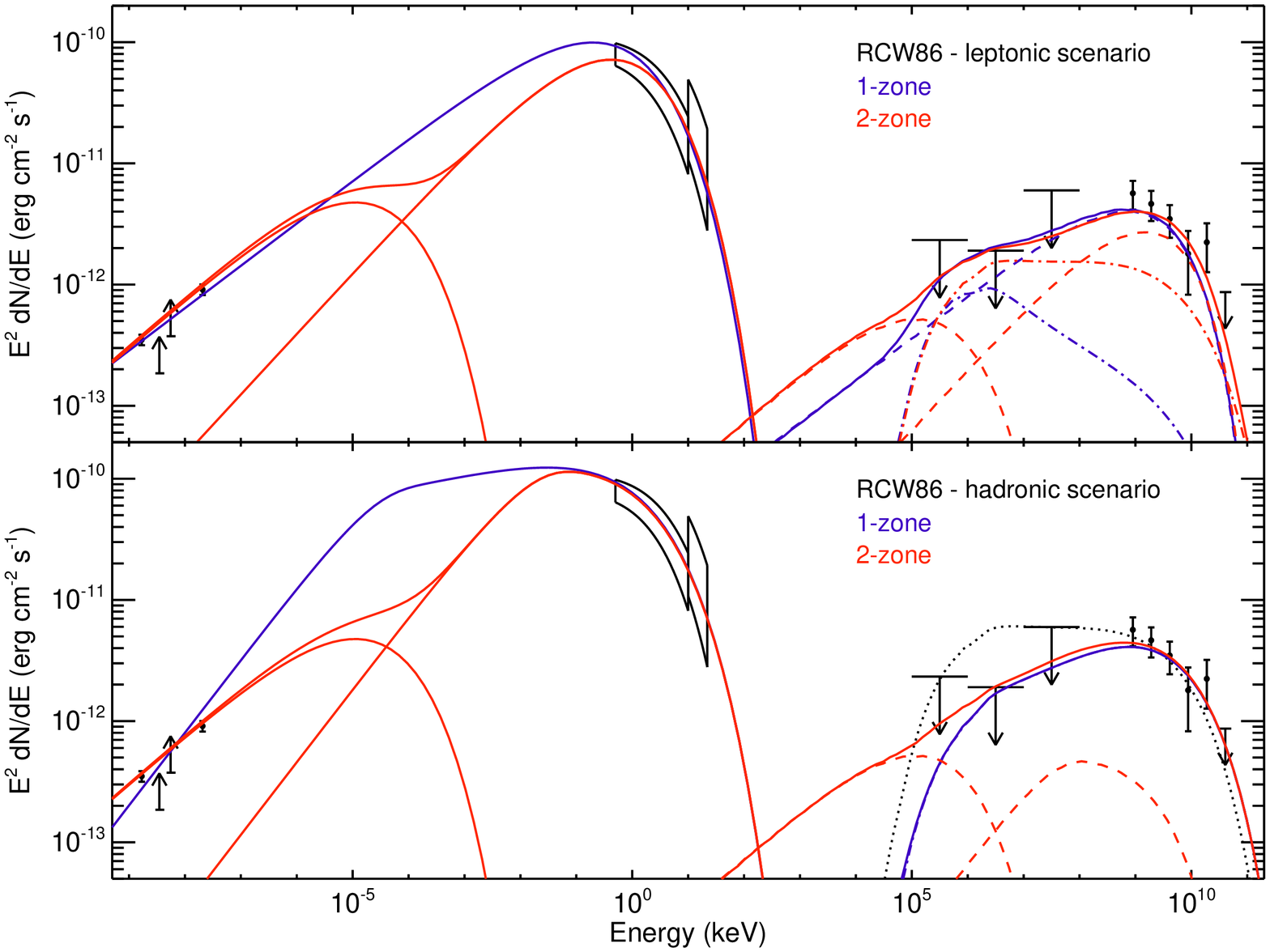}

\caption{Spectral energy distribution of \rcw~with the best-fit leptonic (top) and hadronic (bottom) models. The radio data points 
         are from Molonglo at 408 MHz and \park~at 5 GHz \cite[][D.~A. Green 2011, private communication]{caswell75}, and lower 
         limits from \most~at 843 MHz and \atc~at 1.43 GHz \citep{whiteoak96,dickel01}. \xray~spectra from \asca~and \rxte~were obtained as described in section \ref{xray}. The \fermi-LAT upper limits (99.9\% C.L.) derived in section~\ref{ana} and the 
         \hess~data points in the VHE domain \citep{hess09} are also shown. For each plot, the blue (respectively red) lines denote the total
         broadband emission from the one-zone (respectively two-zone) modeling discussed in section \ref{discu}, with SC, IC (with CMB only)
         and $\pi^0$ spectra shown as solid, dashed, and dash-dotted lines, respectively. The associated best-fit parameters are provided
         in Table \ref{tab2}. Note that in the bottom panel, $\pi^0$ emission is similar for the two models, and equals the total gamma-ray 
         emission in the one-zone model (solid blue line). In addition, the dotted black line shows the (excluded) case of $\pi^0$ emission
         from a E$^{-2}$ hadron spectrum with $\eta_p \sim$ 0.15 d$^2_{2.5}$/\neff$_{{\rm cm-3}}$ and E$_{{\rm max,p}}$ = 100 TeV.
         \label{fig4}}

\end{center}
\end{figure*}

Two mechanisms are commonly proposed to explain the gamma-ray emission in young SNRs: IC scattering off ambient photons (called leptonic 
scenario) and proton-proton interactions (called hadronic scenario). On the basis of a phenomenological approach, we aim to constrain the 
average magnetic field and the total energy in accelerated particles responsible for the broadband nonthermal spectrum of \rcw~shown in 
Figure \ref{fig4}. For this purpose, we modeled the $\pi^0$-decay gamma-ray emission from inelastic interactions of accelerated hadrons 
with ambient gas nuclei, according to the procedure described in \citet{huang07}. The authors calculated the gamma-ray emission from 
the collisions between CR protons and Helium nuclei and the interstellar medium with standard composition. We also computed the exact 
expressions of SC and IC emissions from a parent electron spectrum, as given in \citet{bg70}. Since the contribution to the IC scattering 
from the {\it local} interstellar (optical and infrared) radiation fields (ISRF), at the location of \rcw, is unknown, we considered 
two cases: one with the Cosmic Microwave Background (CMB) as the only source of seed photons for IC scattering, and the other with both the 
CMB and the Galactic ISRFs provided by \citet{porter05}, which comprise the infrared (from dust, at T $\sim$ 35 K) and optical (from stars, 
at T $\sim$ 4600 K) emissions with respective energy densities of 0.66 and 0.94 eV cm$^{-3}$. Nonthermal bremsstrahlung is neglected here 
given the relatively low densities ($\lesssim$ 1 cm$^{-3}$).

We qualitatively fit the free parameters of these leptonic and hadronic scenarios to the available data on \rcw, and discuss the 
corresponding constraints on the parent particle spectrum, energetics, and magnetic field, in terms of the existing estimates. The 
particle spectra are assumed to follow a power-law with an exponential cutoff dN/dE $\propto$ E$^{-\Gamma}$ $\times$ exp(-E/E$_{\rm max}$), 
with the same spectral index for both electrons and hadrons. We note that, in such a case, the resulting SC emission spectrum follows a 
power-law with an exponential cutoff in the form exp(-$(E/E_{\rm cut})^{1/3}$), which is similar to the ZA10 prescription that we used in section 
\ref{xray} to fit the \asca~and \rxte~spectra. We denote the fraction of the total energy injected into accelerated particles 
$\eta_{e,p}$ = W$_{e,p}$/E$_{SN}$, where E$_{SN}$ is the energy of the SN explosion, assumed to be the canonical value of 10$^{51}$ 
erg\footnote{This assumption would be fairly valid in the case of \rcw~is indeed the remnant of a type Ia SN \citep{williams11,yamaguchi11}}.
We also calculate the so-called electron-to-proton ratio $K_{ep}$ at momentum 1 GeV c$^{-1}$, which is compared to the value measured
in cosmic-rays ($K_{ep}$ $\sim$ 10$^{-2}$). In these hadronic and leptonic scenarios, we consider two different models. The first is a one-zone model that assumes that radio, \xrays, GeV, and TeV gamma rays arise from the same emission region with a constant magnetic
field. Since this model cannot account for the morphological difference between radio and \xrays~(see Figure \ref{fig2}), we also consider 
a two-zone model, which exhibits two populations of radio- and \xray-/gamma-ray-emitting particles. The best-fit parameters for these two 
pairs of scenarios are given in Table \ref{tab2}.

\subsection{Leptonic scenario}
\label{lepto}

\subsubsection{One-zone model}
\label{lepto_1zone}

We assume that SC and IC photons are produced by high-energy electrons confined in the same emitting region. Figure \ref{fig4} (top, blue 
lines) shows that the broadband emission of \rcw~can be easily accomodated using an injection spectrum with spectral index $\Gamma$ = 2.3, 
in agreement with the radio measurements, an exponential cutoff at 20--25 TeV, and an average magnetic field of 15--25 $\mu$G, depending 
on the level of ISRFs as seed photons in the calculation of IC emission. This one-zone model also provides an upper limit to the 
energy injected into accelerated hadrons of $\sim 2 \times 10^{49}$ d$^2_{2.5}$/\neff$_{{\rm cm-3}}$ erg, with \neff~the effective
density\footnote{For Bohm diffusion and a compression ratio of 4, $\kappa$ = B$_{up}$/B$_{down}$ $\equiv$ D$_{{\rm down}}$/D$_{{\rm up}}$ = 1/$\sqrt{11}$, as defined in 
Equation \ref{eq1}, with D$_{{\rm down}}$ (resp. D$_{{\rm up}}$) the diffusion coefficient in the downstream (resp. upstream) medium. Given 
that the residence time of particles t scales as 1/D, \neff~=(n$_{{\rm up}}$$\times$t$_{{\rm up}}$ + 
n$_{{\rm down}}$$\times$t$_{{\rm down}}$)/(t$_{{\rm up}}$+t$_{{\rm down}}$) $\sim$ 0.8 n$_{{\rm down}}$. Throughout the discussion, we then 
assume that \neff~equals the post-shock density.}, as depicted by the dash-dotted blue line in Figure \ref{fig4} (top). This estimate is
only slightly sensitive to the assumed E$_{\rm max,p}$, which we conservatively fixed here to 100 TeV. The resulting electron-to-proton 
ratio $K_{ep}$ amounts to $\gtrsim$ 0.05--0.1. 

\begin{table*}[!htb]
  \begin{center}
    \begin{tabular}{|l|c|c|c|c|c|c|}
      \hline\hline

      & \multicolumn{3}{|c|}{Leptonic model}  & \multicolumn{3}{|c|}{Hadronic model}  \\
      & One-zone & \multicolumn{2}{|c|}{Two-zone} & One-zone & \multicolumn{2}{|c|}{Two-zone} \\
      &        &     radio     &     X-ray    &        &      radio    &     X-ray    \\

      \hline

B ($\mu$G)              & 15 (25) & $>$ 10 ($>$ 25) & 15 (25) & 400  & $>$ 10 ($>$ 25) & $>$ 50 \\
$\Gamma$                &   2.3   &      2.2        &   2.0   & 1.8  &      2.2        & 1.8    \\
E$_{\rm break,e}$ (TeV) &   --    &       --        &    --   & 0.04 &       --        & 3.0    \\
E$_{\rm max,e}$ (TeV)   & 25 (20) &   0.2 (0.15)    & 25 (20) & 7    &   0.2 (0.15)    & 20     \\
E$_{\rm max,p}$ (TeV)   &   100   &       --        &   100   & 80   &       --        & 80     \\
$\eta_e$ ($\times$ 10$^{-2}$ d$^{2}_{2.5}$) & 2 (0.9) & $<$ 2.5 ($<$ 0.5) & 0.04 (0.02) & 0.004 & $<$ 2.5 ($<$ 0.5) & 0.006 \\
$\eta_p$ ($\times$ 10$^{-2}$ d$^{2}_{2.5}$/\neff$_{{\rm cm-3}}$) & $<$ 2 & -- & $<$ 4 & 7 & -- & 7 \\
$K_{ep}$                & $>$ 0.1 ($>$ 0.05) & -- & $>$ 0.006 ($>$ 0.003) & 0.002 & -- & 0.001 \\ 

      \hline

    \end{tabular}
    \caption{Best-fit parameters of the leptonic and hadronic scenarios for the one-zone and two-zone models applied to the 
             \rcw~broadband emission (see Fig. \ref{fig4}). Numbers in parentheses correspond to the constraints obtained when 
             considering the Galactic ISRFs from \citet{porter05} in the calculation of IC emission. Note that we do not consider
             any population of accelerated hadrons associated with the radio-SC-emitting electrons in the two-zone models ($K_{ep}$
             is then left undetermined). We also fixed E$_{\rm max,p}$ to 100 TeV in the leptonic scenarios. \label{tab2}}
  \end{center}
\end{table*}

\subsubsection{Two-zone model}
\label{lepto_2zone}

The above estimates rely on the implicit assumption that radio, \xray~and gamma-ray~emission regions occupy the same volume. As shown by 
\citet{rho02} and visible in Figure~\ref{fig2}, this does not seem to be the case for \rcw. Therefore, we consider two different populations 
of {\it radio- and \xray/gamma-ray-}emitting leptons to reproduce the broadband emission of \rcw. In this scenario, we use a harder lepton
spectrum to explain the \xray~and VHE observations, with $\Gamma$ = 2, and a separate electron population to account for the radio emission,
with $\Gamma$ = 2.2. For simplicity, we do not model any putative population of accelerated hadrons associated with the latter leptonic
population, given the poor constraints obtained (see below). Figure \ref{fig4} (top, red lines) shows a reasonable fit of the \xray~and
gamma-ray data assuming an exponential cut-off at 20--25 TeV, and an average magnetic field of 15--25 $\mu$G, which is similar to that of the one-zone
model. The direct consequence of this two-zone model is that it slightly relaxes the constraint on the energy injected into accelerated
particles, owing to a harder spectrum. In this case, the estimated maximal energy injected into accelerated hadrons is $\sim$ 4 $\times$
10$^{49}$ d$^2_{2.5}$/\neff$_{{\rm cm-3}}$ erg, \ie~twice as large as in the one-zone model, and the resulting $K_{ep}$ $\gtrsim$ (3--6)
$\times$ 10$^{-3}$ gets closer to the observed value in CRs. Recently, \citet{williams11} performed a combined IR/X-ray analysis and derived post-shock
densities in the NW and SW regions of the SNR of 2.0 cm$^{-3}$ and 2.4 cm$^{-3}$. If applicable to the whole SNR, these values would imply
a maximal energy injected into hadrons of $\sim 2 \times 10^{49}$ d$^2_{2.5}$ erg, much lower than what \citet{helder09} estimated in the NE 
region, where they evaluated that $> 50$\% of the post-shock pressure is produced by CRs (see below, section \ref{conclu}). However, the 
effective density in this region is known to be much lower than the western part of the shell. For instance, \citet{yamaguchi08} derived a post-shock density from the (weak) thermal \xray~emission of 0.26 $f^{-0.5}$ cm$^{-3}$ ($f$ is the filling factor of this thermal 
component), and \citet{helder09} assumed a density of 0.1 cm$^{-3}$, as derived in \citet{vink06}. As for the radio-SC-emitting electrons, we found that the magnetic field in
this zone must be stronger than 10--25 $\mu$G (depending on the level of Galactic ISRFs). This estimate agrees with the value deduced by 
\citet{vink06} from the width of the filament in the NE region (24 $\pm$ 5 $\mu$G), but barely agrees with the one obtained by \citet{bamba05} for 
the SW filament ($\sim$ 4--12 $\mu$G). \citet{volk05} estimated a stronger magnetic field strength (99$^{+46}_{-26}$ $\mu$G) from the 
thickness of the same SW filamentary structure observed by \citet{rho02}. However, as noted by these authors, this estimate could be reduced 
by a factor of $\sim$ 2, due to the non-spherical geometry, and hence the incomplete de-projection, of this feature, which lies in the 
interior of the remnant. We also note that \citet{arbutina12} found a volume-averaged magnetic field in \rcw~of $\sim$ 70 
$\mu$G, based on modified equipartition calculations applied to shell-type SNRs.

\subsection{Hadronic scenario}
\label{hadro}

\subsubsection{One-zone model}
\label{hadro_1zone}

In the hadronic scenario, gamma rays are predominantly radiated through decays of $\pi^0$ mesons produced in collisions between accelerated 
protons/nuclei and ambient gas. This directly implies that the average magnetic field needs to be much stronger than the previous estimates
to suppress the IC component. In a first step, as performed above, we assume that high-energy electrons and protons are confined in the 
same emitting region. The standard E$^{-2}$-like spectrum with a
fraction of total energy into hadrons of $\eta_p \sim$ 0.15 d$^2_{2.5}$/\neff$_{{\rm cm-3}}$, which would accommodate the \hess~measurements 
\citep[as discussed in][]{hess09}, is clearly ruled out, as shown in Figure \ref{fig4} (bottom, dotted black line). The hard spectral index required is in-between the standard (test-particle) index and the asymptotic limit
of $\Gamma$ = 1.5 predicted by extremely efficient CR acceleration \citep{malkov99,berezhko99}. This is similar to \rxj~\citep{abdo11a} and, to a lesser extent, to RX~J0852.0$-$4622 \citep[aka Vela Junior,][]{tanaka11}. \citet{fang11} modeled the broadband emission of \rcw~based on a semi-analytical solution to the NLDSA theory at SNR shocks. Their predictions
are in conflict with the \fermi-LAT upper limits. It should be noted that the SC spectrum from electrons with the same index $\Gamma = 1.8$ 
is inconsistent with the radio spectral index deduced from the two flux points in Figure \ref{fig4}. To reproduce the radio data from
MOST and Parkes, the magnetic field should be high enough to produce significant energy losses during the lifetime of the remnant ($\sim$
1820 y) and create a break in the SC spectrum. A reasonable fit is obtained by using a total energy injected into accelerated hadrons of
$\sim 7 \times 10^{49}$ d$^2_{2.5}$/\neff$_{{\rm cm-3}}$ erg and an average magnetic field of 400 $\mu$G, with a break in the electron
spectrum at $\sim$ 40 GeV. This magnetic field is much stronger than any estimates published so far on \rcw, as discussed in 
section~\ref{lepto}, and $K_{ep}$ $\sim$ 2 $\times$ 10$^{-3}$ is much lower than that measured in the CR spectrum.  

\subsubsection{Two-zone model}
\label{hadro_2zone}

The constraint on the magnetic field is clearly relaxed in a two-zone model in which the radio emission is produced by a separate electron 
population. In this case, a reasonable fit is shown in Figure \ref{fig4} (bottom, red lines). It assumes a (lower limit on the) magnetic 
field of 50~$\mu$G in the \xray-emitting zone (to suppress the IC component), with an injection spectrum with a power-law index of 
1.8, an energy break at 3 TeV (produced by SC cooling), and an exponential cut-off at 20 TeV. This two-zone model does not modify the 
estimated energy injected into accelerated hadrons, and the spectrum of radio-emitting electrons is also unchanged with respect to the 
two-zone leptonic model (\ref{lepto_2zone}). The resulting $K_{ep}$ is still much lower than 10$^{-2}$. In regard to the spectral shape 
deduced from these hadronic scenarios, we note that \citet{vink06} found that a concave lepton spectrum is favored over a simple 
power-law to explain the broadband SC emission from the NE region, even though they could not distinguish a flattening at high energies 
$\Gamma$ = 2.2 $\rightarrow$ 2 from a flattening $\Gamma$ = 2.2 $\rightarrow$ 1.5. The resulting  $\pi^{0}$-decay emission from such a 
concave hadron spectrum (given that relativistic lepton and hadron indices should be very similar) would certainly mimic an E$^{-1.8}$ 
spectrum at the highest energies, although it is unclear whether the lower-energy part of such a spectrum would still agree with 
the \fermi-LAT upper limits. It is also interesting to note that, in a hadronic scenario, the gamma-ray emission is expected to be 
approximately proportional to the local density. Since the NW and SW regions both exhibit higher densities than the NE region, these 
regions should be brighter in the TeV regime, assuming a uniform CR density along the SNR shell. However, while the \hess~image looks similar to that observed in radio and \xrays, the azimuthal profile shown in \citet{hess09} is well-fit with a constant, without any 
clear variation along the shell. Deeper VHE observations are required to investigate in more detail the gamma-ray morphology of \rcw~and 
help discriminate the origin of the gamma-ray emission.

\section{Conclusion}
\label{conclu}

The \fermi-LAT upper limits in the HE domain derived in this work for the young SNR \rcw~provide strong constraints on the injection 
spectrum of the primary population responsible for the extended VHE emission. A hadronic scenario can only reproduce the multi-wavelength 
data using a hard proton spectrum (spectral index $\Gamma \leq$ 1.8) and a total energy injected into hadrons residing in \rcw~of $\sim$ 7 
$\times$ 10$^{49}$(\neff/1 cm$^{-3}$)$^{-1}$ d$_{2.5}^2$ erg. Given that the one-zone hadronic model suffers from several limitations 
(incompatible radio spectral index, very low $K_{ep}$ and extremely high B-field), only a two-zone model can fulfill all the observational 
constraints, though still with a low $K_{ep}$. In the one- and two-zone leptonic scenarios, the multi-wavelength data can be closely reproduced using electron spectral indices of $\sim$ 2.0--2.3, a total energy injected in electrons of $\sim$ 2 $\times$ 10$^{49}$ erg, 
and a reasonable average magnetic field of 15--25 $\mu$G. In such a case, the most conservative upper limit to the total energy injected 
into hadrons inside \rcw~amounts to $\sim$ 4 $\times$ 10$^{49}$(\neff/1 cm$^{-3}$)$^{-1}$ d$_{2.5}^2$ erg.

These estimates of the total CR energy content in \rcw~can be translated into CR pressure $P_{\rm tot,CR}$ = E$_{\rm tot,CR}$/3\Veff,
where \Veff~is the effective CR volume within the SNR. Assuming that this volume is given by the best-fit parameters of the shell
morphology observed with \hess~\citep[though not statistically significant over a uniform sphere, see][]{hess09}, we obtain \Veff~= 
(6 $\pm$ 2) $\times$ 10$^{59}$ d$_{2.5}^{3}$ cm$^3$, and $P_{\rm tot,CR}$ must be conservatively lower than 3.8$^{+2.0}_{-1.0}$ $\times$ 
10$^{-10}$ (E$_{\rm tot,p}$/7 $\times$ 10$^{49}$ erg) (\neff/0.1 cm$^{-3}$)$^{-1}$ d$_{2.5}^{-1}$ erg cm$^{-3}$. This CR pressure, derived
from the modeling of the broadband nonthermal emission of the {\it entire} SNR (see section \ref{discu}), is very close to the CR pressure
found by \citet{helder09} {\it in the NE region}, and later confirmed by \citet{vink10}, P$_{\rm NE,CR}$ $\gtrsim$ 3.7 $\times$ 10$^{-10}$
(\neff/0.1 cm$^{-3}$) (kT/2.3 keV) erg cm$^{-3}$. If we were to apply the same comparison for the other regions along the \rcw~shell, which 
all exhibit higher plasma densities \citep[0.5--2 cm$^{-3}$, see][]{yamaguchi11,williams11}, our upper limit to $P_{\rm tot,CR}$ would most 
likely be inconsistent with the \citet{helder09} estimate regarding the fractional CR pressure. However, the large uncertainties in the
effective CR volume in the above calculations, among others, prevent us from drawing firm conclusions about the acceleration efficiency in \rcw.


\begin{acknowledgements}

The \fermi~LAT Collaboration acknowledges generous ongoing
support from a number of agencies and institutes that have supported
both the development and the operation of the LAT as well as scientific data
analysis. These include the National Aeronautics and Space Administration and
the Department of Energy in the United States, the Commissariat \`a l'Energie
Atomique and the Centre National de la Recherche Scientifique / Institut
National de Physique Nucl\'eaire et de Physique des Particules in France, the
Agenzia Spaziale Italiana, the Istituto Nazionale di Fisica Nucleare, and the
Istituto Nazionale di Astrofisica in Italy, the Ministry of Education, Culture,
Sports, Science and Technology (MEXT), High Energy Accelerator Research
Organization (KEK) and Japan Aerospace Exploration Agency (JAXA) in Japan,
and the K. A. Wallenberg Foundation and the Swedish National Space Board
in Sweden. Additional support for science analysis during the operations phase
from the following agencies is also gratefully acknowledged: the Instituto
Nazionale di Astrofisica in Italy and the Centre National d'Etudes Spatiales in
France.

We thank D.~A.~Green for helpful advice regarding radio measurements, and
A. Marcowith for stimulating discussions about the modeling of the \rcw~broadband
emission. 

\end{acknowledgements}



\end{document}